# OJ 287: NEW TESTING GROUND FOR GENERAL RELATIVITY AND BEYOND

C Sivaram

Indian Institute of Astrophysics, Bangalore

Abstract: The supermassive short period black hole binary OJ287 is discussed as a new precision testing ground for general relativity and alternate gravity theories. Like in the case of binary pulsars, the relativistic gravity effects are considerably larger than in the solar system. For instance the observed orbital precession is forty degrees per period. The gravitational radiation energy losses are comparable to typical blazar electromagnetic radiation emission and it is about ten orders larger than that of the binary pulsar with significant orbit shrinking already apparent in the light curves. This already tests Einstein gravity to a few percent for objects at cosmological distances. Constraints on alternate gravity theories as well as possible detection of long term effects of dark matter and dark energy on this system are described.

For more than fifty years after Einstein proposed the general theory of relativity in 1915, observational tests to verify some of the predictions were confined to within the solar system; where the effects are quite small. This situation changed with the discovery of the binary pulsar in 1975 where the relativistic periastron shift was more than four degrees per year, a whopping thirty thousand times more than the paltry well known correction of 43 arc seconds/century for mercury.[1, 2] The recently discovered 2.4 hour period binary pulsar has a periastron shift of sixteen degrees per year![3] Other relativistic effects like the time delay of the signals and time dilation and frequency shifts are also much larger for these binary systems.

Moreover due to gravitational radiation losses (as predicted by the quadrupole formula of GR), the binary pulsar components are approaching each other with the corresponding observed shortening of the periods. The 2.4 hour binary pulsar is expected to merge in about a million years. The monitoring of the Hulse-Taylor pulsar for the past thirty years has enabled general relativity to be verified to about 0.1 percent precision, putting tight constraints on alternative gravitational theories.[4]

However gravitational waves have yet to be directly detected with the ongoing LIGO detectors well on the search[5] and the space borne LISA due to go up in the next decade. The gravity B probe is currently testing smaller effects such as geodetic and Lense-Thirring precessions due to the earth's rotation. The latter is only of the order of 0.1 arcsec/ year! For binary neutron stars, the geodetic precession is ~0.5 deg/yr and has been observed.[6]

The effects of GR and strong gravitational field effects should be even more important for binary black holes, as these are more compact objects. Indeed the expected gravitational waves from inspiraling binary black holes are a stronger motivation for the LISA probe which can detect such mergers even a gigaparsec away! The detectable redshift from line x-rays can, it is claimed, even determine the rotation of the black hole by probing the innermost circular orbit around the accretion disc.[7]

Unlike the binary pulsars we have so far not had a similar black hole binary which could provide precision tests of relativistic effects. Recently the long studied massive binary black hole system OJ287 has provided this opportunity.[8]

This system consists of two orbiting supermassive black holes about a gigaparsec away and flares up every 12 years. Its optical light curve has been studied for more than hundred years (from 1891).[8] The periodic optical outbursts occur soon after the secondary black hole has impacted on the accretion disc of the primary.

Information on the exact timings of six such occurrences is enough to produce a unique solution of the orbit and after the 2005 major outburst the full orbit determination became possible.

Radiation from the disc flares up to several thousand times in two bursts just over a year apart, as every twelve years, the smaller black hole passes through the accretion disc twice, stirring it up to produce the brightness peaks. New flares appeared in 1994, 1995 and November 2005.

The mass of the primary black hole has been deduced at 18 billion solar mass, a new record for a supermassive black hole (SMBH), six times larger than the 3 billion solar mass estimated for the SMBH in the giant elliptical M87 nucleus. The secondary black hole of OJ287 has a mass of $10^8$ solar mass.[8]

For the orbital period of 12 years, the orbit separation is $\sim 10^4$ astronomical units (A. U). This gives a GR recession rate of the binary orbit of about forty degrees per period, which is confirmed by observations. This is comparable with that of the Hulse-Taylor binary pulsar!

The gravitational time delay (the Shapiro effect) given by:
$$\frac{4GM}{c^3} \times \log \text{factor} \sim 10 \text{ days} \qquad \ldots (1)$$
for this system.

Thus without the effect of this space time curvature on the propagation of light, the flare observed on 13 September 2007, would have occurred <u>10 days earlier!</u>

The 13 September flare was observed by 30 astronomers at the 3.5m telescope at Calar Alto observatory and the 2.5m Nordic optical telescope, Canary Island.[8] The mass implied by the time delay agrees with that estimated from the observed relativistic precession![8]

$$\Delta\phi_{rel} = \frac{6\pi G(M_1 + M_2)}{ac^2(1-e^2)} = 39\deg/\,period!(e=0.7) \qquad \ldots (2)$$

It is exciting that here we have a supermassive black hole binary where the observable effects of gravitational radiation are far more dramatic than that of the well studied binary pulsar. Here also the orbit shrinks due to gravitational radiation emission. The orbital velocity is:

$$v_{Orb} \sim 0.1c \sim 30,000 km/s \qquad \ldots (3)$$

For the binary pulsar, it is a mere *300km/s*! As the rate of emission of gravitational waves goes as $\left(v_{Orb}/c\right)^5$, we would expect this system to shrink considerably more and merge on a much shorter time scale than the binary pulsar.

The energy loss due to gravitational radiation is given as:[9]

$$\dot{E}_G = \frac{32G}{5c^5} M^2 R^4 \omega^6 f(e) \sim 3\times 10^{41} Watts! \qquad \ldots (4)$$

(*f(e)* is a function of orbital eccentricity)

This energy loss is the highest known gravitational wave energy emission for any object and is comparable to the electromagnetic energy emission for typical quasars![10]

The orbital shrinking rate can be estimated from:

$$\dot{E}_G \approx \frac{GM_1 M_2}{a^2} \dot{R} \text{ (etc.)} \qquad \ldots (5)$$

From (4), this implies that the orbit shrinks by 40km/s, so that the black holes would merge on a time scale of less than ten thousand years! Due to the orbit shrinking, the period shortens by (as estimated from $\dot{P} \approx 6\pi R^2/Gm_P \, \dot{R} f(e)$) 0.04 S/sec, so that period between the bursts come <u>20 days sooner</u>! So without the gravitational radiation energy loss the flare would have occurred 20 days after 13 September 2007.

So both the effects observed due to gravitational wave energy loss and time delay test general relativity to within <u>five percent</u>, for such a supermassive binary black hole, 1 Gpc away!

The gravitational wave energy released when the merger occurs would be $\sim 10^{55}$ Joules. Even at Gpc distances the strain on a gravitational wave detector would be:

$$h \sim GE/c^4 r \sim 10^{-16} \qquad \ldots (6)$$

(Typical frequency ~5-10 microhertz)

This would be a very 'bright' signal for LISA (with a threshold of ~$10^{-23}$)[11]

The geodesic precession for this system is:

$$\dot{\theta}_{geod} \approx \frac{3}{2} \frac{GM}{R^2} \left(\frac{GM}{R}\right)^{1/2} \sim 0.5 \deg/year \qquad \ldots (7)$$

This is comparable with that of the binary neutron star systems. The Thomas precession $a \times v/c^2 \sim 0.2 \deg/year$. Both these effects are now being measured for earth orbit by the Gravity B probe.[12] Further monitoring of OJ287 and complimentary results from GPB could soon put constraints on alternate gravity theories.

MOND would not be testable as far as this system is concerned as the field strengths are large.[13] Bekenstein has proposed a relativistic basis for MOND which is a tensor-vector-scalar (TeVeS) theory, as an alternative to GR.[14] As these types of theories involve a long range vector component, they could cause energy loss of the binary by dipole radiation.[15,16] From the binary and millisecond already tight limits on the strength of such a component has been placed.[15,16] OJ287 would place a constraint on the relativistic strength $\alpha_v$ of such a coupling (as compared to Newtonian force) of $\alpha_v < (v/c)^2 \times 10^{-2} \approx 10^{-5}$.

This is not so stringent as the pulsar limits or from lunar laser ranging which implies $\alpha_v < 10^{-10}$.[17]

However further monitoring and observation of other such systems has the potential to improve the limits considerably. Possible manifestations of the effects of dark energy (DE) and dark matter (DM) on the dynamics of such systems could be also estimated.[18] DE and DM together constitute more than ninety-five percent of the matter in the universe.[19]

The DE currently is almost consistent with a cosmological constant term. For systems of large separation it is more important.

It is interesting to estimate the DE contribution to the relativistic precession of OJ287. It is given by a term of the form:[18]

$$\Delta\theta \approx 2\pi\left(1 - \frac{3\lambda r_0^3}{2GM}\right) \approx 3\pi\Lambda r_0^3 c^2/GM \qquad \ldots (8)$$

For the observed cosmic DE density, this works out to a mere picoarcsec per period for OJ287. However it increases as cube of separation. So for a system of ten times the separation and one tenth the mass, it is a more respectable one microarcsec![18]

The DM contribution, considering accretion by SMBH is significantly larger, i.e., ~0.1 milliarcsec/year. A HipparCOS type probe or VLBI could in principle detect this.[20] Again DM could cause a period change of $10^{-4}s/sec$ and DE a change of $10^{-9}s/sec$[20] for this system. These are atomic clock type precisions!

Finally it has been suggested[21], that Intra-Day-Variability (IDV) of objects like blazars could help differentiate between black holes and worm holes. As none of the periodicities predicted for worm holes (for corresponding masses) (like the radial and gyrofrequencies associated with the worm hole neck) are present in OJ287, such scenarios are ruled out. In short this massive black hole binary system is a new unique precision testing ground for various relativistic gravity effects over cosmological scales.